\documentclass[aps,pra,amssymb,twocolumn,superscriptaddress,showpacs,floatfix]{revtex4}
\usepackage{psfrag} \usepackage{bm}
\usepackage{color} \usepackage[english]{babel}
\usepackage{amsmath} \usepackage{graphicx}
\usepackage{amssymb} \usepackage{amsfonts} \usepackage{subfigure}
\newcommand{\s}{\mbox{\boldmath$\sigma$}}
\newcommand{\ks}{\mbox{\scriptsize \boldmath $\sigma$}}

\newcommand{\1}{\leavevmode{\rm 1\ifmmode\mkern  -4.8mu\else\kern -.3em\fi I}}
\newcommand{\ket}[1]{\vert #1 \rangle}

\newcommand{\braket}[2]{\langle #1 \vert #2 \rangle}

\newcommand{\ketbra}[2]{\vert #1 \rangle \! \langle #2 \vert}

\newcommand{\Tr}[1]{\textrm{Tr}\,\! #1}

 \newcommand{\beq}{\begin{equation}}
\newcommand{\eeq}{\end{equation}} \newcommand{\barr}{\begin{eqnarray}}
\newcommand{\earr}{\end{eqnarray}} \newcommand{\andy}[1]{ }

 \def\ket#1{| #1 \rangle}

\renewcommand{\Re}{\mathop{\text{Re}}\nolimits}

\begin{document}
\title{Optimal quantum estimation in spin systems at criticality}
\author{Carmen Invernizzi}
\affiliation{Dipartimento di Fisica, Universit\`a di Milano, I-20133 Milano, Italia }
\author{Michael Korbman}
\affiliation{Dipartimento di Fisica, Universit\`a di Milano, I-20133 Milano, Italia }
\author{Lorenzo Campos Venuti}
 \affiliation{ISI Foundation, I-10133 Torino, Italia}
\author{Matteo G. A. Paris } 
\affiliation{Dipartimento di Fisica, Universit\`a di Milano, I-20133 Milano, Italia }
\affiliation{ISI Foundation, I-10133 Torino, Italia}
\affiliation{CNISM, UdR Milano, I-20133 Milano, Italia}
\date{\today}
\begin{abstract}
It is a general fact that the coupling constant of an interacting
many-body Hamiltonian do not correspond to any observable and one has to
infer its value by an indirect measurement.  For this purpose, quantum
systems at criticality can be considered as a resource to improve the
ultimate quantum limits to precision of the estimation procedure. In
this paper, we consider the one-dimensional quantum Ising model 
as a paradigmatic example of many-body system exhibiting
criticality, and derive the optimal quantum estimator of the 
coupling constant varying size and temperature. 
We find the optimal external field, which maximizes
the quantum Fisher information of the coupling constant, both for few spins
and in the thermodynamic limit, and show that at the critical point a
precision improvement of order $L$ is achieved. We also show that the 
measurement of the total magnetization provides optimal estimation for 
couplings larger than a threshold value, which itself decreases with 
temperature.
\end{abstract}
\pacs{03.65.Ta, 05.70.Jk, 06.20.-f}
\maketitle
\section{Introduction}
\label{sec:Introduction}
Acquiring information about a physical system involves observations and
measurements, whose results are subjected to fluctuations, and one would
like to eliminate or at least to minimize the corresponding errors.
However, the precision of any measurement procedure is bounded by
fundamental law of statistics and quantum mechanics, and in order to
optimally estimate the value of some parameter, one has to exploit the
tools provided by quantum estimation theory (QET) \cite{Helstrom}. 
\par
As a matter of fact, many quantities of interest do not correspond to
quantum observables. Relevant examples are given by the entanglement or 
the purity of a quantum state \cite{mg} or the coupling constant of 
an interacting Hamiltonian. In these situations one needs to infer the 
value of the parameter through indirect measurements. For many-body quantum 
systems, changing the coupling constant drives the system into different 
phases and, in turn, this may be used to estimate the coupling itself. 
In particular, close to critical points, quantum states
belonging to different phases should be distinguished more effectively
than states belonging to the same phase \cite{pzPRE06, HQZ, mcpg, mcri,
lcvPRL07, pzpgmcPRL07,pzPRA07}.  Distinguishability is usually quantified by
fidelity between quantum states, {\em i.e.} overlap between ground state 
wave functions. In turn, the fidelity approach to 
quantum phase transitions (QPT) has recently attracted much attention 
\cite{pzPRE06, ZZL} since, differently from bipartite
entanglement measure approach \cite{letteraturaSpinChains}, it considers
the system as a whole, without resorting to bipartitions. 
In estimating the value of a parameter, one is led to define the 
Fisher information which represents an infinitesimal distance among
probability distributions, and gives the ultimate precision attainable 
by an estimator via the Cramer-Rao theorem. Its quantum counterpart, the
quantum Fisher information (QFI), is related to the degree of
statistical distinguishability of a quantum state from its neighbors 
and, in fact, it turns out to be proportional to 
\textit{Bures metric} between quantum states 
\cite{bur69,uhl76,woo81,jos94,QCR,QCR2,bro9X,nag00}.
\par
As noticed in \cite{pzmp} one can exploit the geometrical theory of
quantum estimation to derive the ultimate quantum bounds to the
precision of any estimation procedure, and the fidelity approach to QPTs
to find working regimes achieving those bounds.  Indeed, precision may
be largely enhanced at the critical points in comparison to the regular
ones.  Here we show that the general idea advocated in \cite{pzmp} can
be successfully implemented in systems of interest for quantum
information processing.  To this aim we address a paradigmatic example
of many-body system exhibiting a (zero temperature) QPT: the
one-dimensional Ising model with a transverse magnetic field. 
\par
In most physical situations, some parameters of the Hamiltonian, {\em
e.g.} the coupling constant, are unaccessible, whereas others may be
tuned with reasonable control by the experimenter ({\em e.g.} external
field). Therefore, the idea is to tune the controllable parameters in
order to maximize the QFI and thus the distinguishability and the
estimation precision. In doing this we consider the system both at zero
and finite temperature, and fully exploit QET to derive the optimal
quantum measurement for the unobservable coupling constant in terms of
the symmetric logarithmic derivative. In the thermodynamic limit we
find that optimal estimation is achieved tuning the field at the
critical value, in accordance with \cite{pzmp}, whereas at finite size
$L$, the request of maximum QFI defines a pseudo-critical point which
scales to the proper critical point as $L$ goes to infinity. In turn, a
precision improvement of order $L$ may be achieved with respect to the
non critical case. 
\par
The optimal measurement arising from the present QET approach
may be not achievable with current technology. Therefore, having 
in mind a practical implementation, we consider estimators
based on feasible detection schemes, and show, for systems of few 
spins, that the measurement of the total magnetization allows for 
estimation of the coupling constant with precision at the 
ultimate quantum level. 
\par
The paper is structured as follows: In Section \ref{def} we briefly
review some concepts of QET, introduce the symmetric logarithmic
derivative and illustrate the quantum Cramer-Rao bound. We also review
the notion of distance for the quantum Ising model. In Section
\ref{Tzero} we derive the ultimate quantum limits to the precision of
coupling constant estimation at zero temperature, both for the case of
few spins and then in the thermodynamical limit. In Section \ref{Tfinite} 
we analyze the effects of temperature and derive the scaling properties of QFI. 
In Section
\ref{magnetization} we address the measurement of total magnetization 
as estimator of the Hamiltonian parameter and show its optimality.
Section \ref{conclusion} closes the paper with some concluding remarks.
\section{Preliminaries}\label{def}
In this section we recall the basic concepts of QET and the metric 
approach to quantum criticality, specializing them to the one-dimensional 
Ising model in transverse field.  
\subsection{Quantum Estimation Theory}
An estimation problem consists in inferring the value of a parameter
$\lambda$ by measuring a related quantity $X$. The solution of the
problem amounts to find an estimator $\hat{\lambda}\equiv \hat\lambda
(x_1, x_2, \ldots)$, {\em i.e.} a real function of the measurements
outcomes $\{x_k\}$ to the parameters space.  Classically, the variance
$\hbox{Var}(\lambda) = E [\hat\lambda^2]  - E [\hat\lambda]^2$ of any 
unbiased estimator satisfies the Cramer-Rao
theorem 
$$\hbox{Var}(\lambda)\geq \frac{1}{M F(\lambda)}\:,$$
which establishes a lower bound on variance in terms of the number of independent
measurements $M$ and the Fisher Information (FI) $F(\lambda) = E
\left[\left( \partial_\lambda \log p(x|\lambda)\right)^2\right]$ {\em i.e.}
\begin{equation}\label{fisherclassica} F(\lambda) =\sum_x p(x |
\lambda) \left [\partial_\lambda \log p(x | \lambda)\right ]^2\:,
\end{equation} 
$p(x | \lambda)$ being the conditional probability of
obtaining the value $x$ when the parameter has the value $\lambda$.
When quantum systems are involved 
$p(x | \lambda) =\hbox{Tr}\left[\varrho_\lambda\,P_x\right]$, $\{P_x\}$
being the probability operator-valued measure (POVM) describing the
measurement. A quantum estimation problem thus corresponds to 
a quantum statistical model, {\em i.e.} a set of quantum states 
$\rho_\lambda$ labeled by the parameter of interest, with the mapping
$\lambda\to\rho_\lambda$ providing a coordinate system.
Upon introducing the Symmetric Logarithmic Derivative (SLD) $\Lambda_\lambda$
as the set of operators satisfying the equation 
\begin{equation}\label{SLD}
\partial_\lambda\rho _\lambda=\frac{1}{2} \Big [ \Lambda_\lambda \rho _\lambda 
+ \rho _\lambda\Lambda_\lambda \Big ],
\end{equation}
we can rewrite the FI as 
\begin{equation} 
F(\lambda) =\sum_x \frac{\Re \left (\Tr \left [ 
\rho _\lambda\Lambda_\lambda P_x\right ]\right )^2}{\Tr 
\left [ \rho _\lambda P_x\right ]}\:.
\end{equation}
Then one can prove \cite{QCR, QCR2} that $F (\lambda)$ is upper bounded by the 
Quantum Fisher Information 
\begin{equation}
F(\lambda) \leq G(\lambda)\equiv \Tr \left [ \rho_\lambda
\Lambda_\lambda^2\right ]\:.
\label{QFI}
\end{equation}
In turn, the ultimate limit to precision is given by the quantum
Cramer-Rao theorem (QCR) 
$$\hbox{Var}(\lambda)\geq \frac{1}{M G(\lambda)}\:,$$
which provides a measurement-independent lower bound for the variance 
which is attainable upon measuring a POVM built with the
eigenprojectors of the SLD.
\par
In the following we will consider the quantum statistical model defined 
by the set of Gibbs thermal states $\rho_\lambda = Z^{-1} 
e^{-\beta H(\lambda)}$, ($Z= \Tr[e^{-\beta H(\lambda)}]$) associated
with a family $H(\lambda)$ of many-body Hamiltonians  where
$\lambda$ is the coupling constant we wish to estimate.
The relevant observation at this point is that the Bures distance 
$d^2_B (\rho, \rho')$ between quantum states at nearby points in 
parameter space  may be written as 
\begin{equation}
ds^2_\lambda \equiv d^2_B (\rho_\lambda,\rho_{\lambda+d\lambda})  
= g_{\lambda} d\lambda^2 = \frac14 G(\lambda) d\lambda^2 
\end{equation}
where $G(\lambda)$ is the QFI defined in Eq. (\ref{QFI}) and $g_\lambda$ is 
Bures metric tensor given by  \cite{SZ2003}
\begin{equation}
g_\lambda = \frac12 \sum_{nm} \frac{\left|\langle \psi_m| \partial_\lambda
\varrho_\lambda | \psi_n\rangle\right|^2}{p_n+ p_m}
\label{HH}\:,
\end{equation}
$|\psi_n\rangle$ being the eigenvectors of $\rho_\lambda= \sum_n p_n
|\psi_n\rangle\langle \psi_n |$.
In other words, maxima of the Bures metric, {\em e.g.}
the divergence occurring at QPTs \cite{pzpgmcPRL07}, correspond
to optimal estimation working regimes. In the following we will
systematically seek for maxima of Bures metric (QFI). In the
thermodynamic limit those occur at the critical points \cite{pzpgmcPRL07},
whereas at finite size the maxima of the QFI define pseudo-critical 
points which scale to the actual critical points as the size goes to 
infinity.
\subsection{Quantum Ising model}
We are interested in systems which undergo a zero-temperature quantum
phase transition and consider a paradigmatic example, the one-dimensional 
quantum Ising model of size $L$ with transverse field. The model is defined by 
the Hamiltonian
\begin{equation}\label{HIsing}
H= - J\sum_{k=1}^L \sigma_k^x\sigma_{k+1}^x - h\sum_{k=1}^L \sigma_k^z,
\end{equation}
where the $\sigma^\alpha_i$, are Pauli operators and we assume periodic 
boundary conditions $\sigma_{L+1}^x = \sigma_{1}^x$ unless stated otherwise.  
As the temperature and the field $h$ are varied one may identify different 
physical regions.  At zero temperature, the system undergoes a QPT for $h=J$.  
For $h<J$ the system is in an ordered phase whereas for $h>J$ the field 
dominates, and the system is in a paramagnetic state. For temperature $T\ll 
\Delta$, $\Delta= \left| J-h\right|$ the system behaves quasi-classically, 
whereas for $T\gg\Delta$ quantum effects dominate. The Hamiltonian (\ref{HIsing}) 
can be exactly diagonalized by a Bogoliubov transformation, leading to
\begin{equation}
H=\sum_{k>0} \Lambda_k \left ( \eta_k^\dagger\eta_k -1 \right) \, ,
\label{eq:ising_diag}
\end{equation}
where $\Lambda_k $ denotes the one particle energies and $\eta_k$ the
fermion annihilation operator,
$\Lambda_{k}=\sqrt{\epsilon_{k}^{2}+\Delta_{k}^{2}}$, $\Delta_k = J\sin
(k)$, $\epsilon_k =(J\cos (k) +h)$. 
Strictly speaking, Eq.~(\ref{eq:ising_diag}) holds in the sector with even
number of fermions. In this case, periodic boundary conditions on the
spins induce antiperiodic BC's on the fermions and the momenta satisfy
$k=\frac{(2n+1)\pi}{L}$. In the sector with odd number of particles,
instead, one has $k=\frac{(2n)\pi}{L}$ and one must carefully treat
excitations at $k=0$ and $k=\pi$.  In any case, the ground state of
(\ref{HIsing}) belongs to the even sector so that, at zero temperature
we can use Eq.~(\ref{eq:ising_diag}) for any finite $L$.  At positive
temperature we will be primarily interested in large system sizes and
therefore we can neglect boundary terms in the Hamiltonian and use
Eq.~(\ref{eq:ising_diag}) in the whole Fock space. For small $L$ we will
diagonalize explicitly the Hamiltonian (\ref{HIsing}), without
resorting to (\ref{eq:ising_diag}).
\par
The QFI for the parameter $J$ may be evaluated starting from Eq.
(\ref{HH}) arriving at 
\begin{equation}\label{g_JJ}
G_J =\sum_n \frac{(\partial_J p_n) ^2}{p_n} + 2\sum_{n\neq
m}\left|\braket{\psi_n}{\partial_J \psi_m}\right|^2\frac{(p_n -p_m)^2}{p_n+p_m},
\end{equation}
which, given $E_j =\sum_k n_k\Lambda_k$, where the $n_k$'s are the 
fermion occupation numbers, may be written as \cite{pzPRA07}
\begin{eqnarray}\label{g-QFI}
G_J (J,h,\beta) &= &\frac{\beta^{2}}{4}\sum_{k}
\frac{\left(\partial_{J}\Lambda_{k}\right)^{2}}{\cosh^{2}
\left(\beta\Lambda_{k}/2\right)} \nonumber \\
&+ & \sum_{k}\frac{\cosh\left(\beta\Lambda_{k}
\right)-1}{\cosh\left(\beta\Lambda_{k}\right)}
\left(\partial_{J}\vartheta_{k}\right)^{2}.
\end{eqnarray}
where $\vartheta_k = \hbox{arctan}\frac{\epsilon_k}{\Delta_k}$. 
Since the QFI is proportional to the Bures metric one may exploit 
the results derived for the Bures metric, which we recall here:
\begin{itemize} 
\item At zero temperature, in the off-critical region (the 
{\em thermodynamic limit}) $L \gg \xi$, where $\xi$ is the 
system correlation length,  the Bures metric behaves as 
$g_\lambda \sim L \left | \lambda - \lambda_c \right 
|^{-\Delta_g}$ close to the critical point. Here $L$ is the 
size of the system and $\Delta_g$ is related to the 
critical exponents of the transition, which turns out to be one for
our system \cite{lcvPRL07,pzpgmcPRL07}.
\item In the quasi-critical region $\xi \gg L$, $g_\lambda$ scales 
as $g_\lambda \sim L^\alpha$ with $\alpha=1+\Delta_g/\nu$ where 
$\nu$ is the correlation length critical exponent {\em i.e.} $\xi 
\sim |\lambda - \lambda_c|^{-\nu}$. It turns out that $\alpha=2$ 
quasi-free fermionic models, as the one considered here.
\item At regular points ({\em i.e.} not-critical) the Bures 
metric is extensive {\em i.e.} $g_\lambda \sim L$.
\item As the temperature is turned on, as long as it is small 
but larger than the energy gap of the system,  quantum-critical  
effects dominate. In this region  $T\gg\Delta$, thermodynamic 
quantities scales algebraically with the temperature and one 
has $g_\lambda \sim T^{-\beta}$, with $\beta >0$. For the 
Ising model $\beta =1$ \cite{pzPRA07}.
\end{itemize}
In the following, we will exploit the dramatic increase of the QFI 
that one experiences in the critical regions, to improve the ultimate 
quantum limit achievable in the estimation of the coupling parameter.
\section{QET at zero temperature}\label{Tzero}
In this section we begin to test the idea of estimating the coupling constant 
$J$ of the Ising model by finding the maximum of QFI at zero
temperature, where the system is in the ground state. At first we consider 
systems made of few spins and then we address the 
thermodynamic limit.
\subsection{Small $L$ }
We start with the case of $L=2, 3$ and $4$ in Eq.(\ref{HIsing}). 
The QFI  is obtained from Eq.~ (\ref{g_JJ}) by explicit diagonalization 
of the Ising Hamiltonian where $p_n =e^{-\beta E_n}/Z$, $E_n$ and 
$\ket{\psi_n}$ are the eigenvalues and eigenvectors of $H$. For example, 
for  $L=2$ we have $E_n = \pm 2 J, \pm2 \sqrt{J^2 + h^2}$ and $Z= 2\cosh (2\beta J) 
+ 2\cosh (2\beta\sqrt{J^2+h^2})$. For $T=0$  one gets  
\begin{eqnarray} 
G_J (J,h,0)&=&\frac{h^2}{(h^2 + J^2)^2}, \quad L =2 \nonumber \\
G_J(J,h,0)&=&\frac{3 h^2}{4(h^2 - hJ + J^2)^2}, \quad L =3 \nonumber \\
G_J(J,h,0)&=&\frac{h^2 (h^4 + 4h^2J^2 + J^4)}{(h^4+J^4)^2}, \quad L =4 
\label{gtz} \, .
\end{eqnarray}
Maxima of $G_J$ are obtained for $h^\ast=J$ for $L=2,3,4$. Actually, this is
true for any $L$, see also the next Section, and the pseudo-critical point 
$h^\ast$ which maximizes $G_J$, turns out to be independent of $L$ and 
equal to the true critical point, $h_c= J, \, \forall L$. At its maximum 
$G_J$ goes like $1/J^2$ and the ultimate lower bound to precision
(variance) of any quantum estimator of $J$ scales as $J^2$.
\subsection{Large $L$ }\label{Lspin}
In the following we discuss the QFI for a system of size $L$. 
We analyze the behavior of $G_J$  near the critical region at $T=0$.
Taking the limit $T\to 0$ in Eq.(\ref{g-QFI}), the classical elements of 
the Bures metric, which depends only on thermal fluctuations, vanishes due 
to the factor of $(\cosh (\beta\Lambda_k/2))^{-2}$. Therefore, at zero temperature, 
only the nonclassical part of Eq.(\ref{g-QFI}) survives and one obtains
\begin{equation}\label{G_JTzero}
G_J = \sum_{k } (\partial_J\vartheta_k)^2,
\end{equation}
where
$\partial_J \vartheta_k =\frac{1}{1+ (\Delta_k /\epsilon_k )^2} 
(\partial_J \frac{\Delta_k}{\epsilon_k})= \frac{- h\sin k}{\Lambda_k^2}$. 
Since we are in the ground state, the allowed quasi-momenta are $k = 
\frac{(2 n+1)\pi}{L}$ with $n=0,\ldots, L/2-1$.  Explicitly we have
\begin{eqnarray}\label{G_J}
G_J &=&\sum_k \frac{h^2\sin (k)^2}{\Lambda_k^4}. 
\end{eqnarray}
We are interested in the behavior of the QFI in the quasi-critical region 
$\xi \gg L$. In the Ising model $\nu=1$ so the critical region is described 
by small values of the scaling variable $z \equiv L (h-J) \simeq L/\xi$, that 
is $z\approx 0$. Conversely the off-critical region is given by $z\rightarrow\infty$.
We substitute $h =  J+ z/L$  in Eq.(\ref{G_J}) and expand around $z=0$ to obtain 
the scaling of $G_J$ in the quasi-critical regime
\begin{equation}
G_J = \sum_{k_n}\frac{(J+\frac{z}{L})^2\sin^2 (k_n)}{[\frac{z^2}{L^2} 
+ 4J(J+\frac{z}{L})\sin ^2(k_n/2)]^2} \equiv \sum_{k_n} f_{k_n}(z) \, .
\end{equation}
Since $\partial_z f(0)=0$,  the maximum of $G_J$ is always at $z=0$ for all 
values of $L$, in turn, the pseudo-critical point is $h_L^\ast = J = h_c$ $\forall 
L$.  As already noticed previously, the statement $h_L^\ast  = h_c$ is peculiar 
to this particular  situation. For instance, introducing an anisotropy $\gamma$ so 
as to turn the Ising model into the anisotropic $XY$ model, the pseudo-critical 
point gets shifted and one recovers the general situation  $h_L^\ast =  h_c + 
O\left( L^{-\theta}\right)$. 
Going to second order one obtains
\begin{eqnarray}\label{sum_thetak}
\sum_k (\partial_J \theta_k)^2  &=&  \sum_{k_n}\frac{1}{4 J^2}\cot^2(k_n/2) 
\Big ( 1 -\frac{z^2}{2 J^2 L^2}\times \nonumber
\\ &&\frac{1}{\sin^2 (k_n/2)} \Big )+ O\left (z^3\right ) \, .
\end{eqnarray}
Using Euler-Maclaurin formula \cite{EMC} 
we get
\begin{equation}
G_J = L^2 \left(\frac{1}{8 J^2}  - \frac{z^2}{384 J^4} \right)  
-\frac{L}{8 J^2} + O(L^0).
\end{equation}
This shows explicitly that at $h=J$  the Fisher information 
has a maximum and there it behaves as
\begin{equation}
G_J (L, T=0, h^\ast=J)\simeq \frac{L^2}{8 J^2} + O(L).
\end{equation}
We observe that superextensive behavior of the QFI in the quasi-critical
region around the  QPT, $G_J \sim L^2 $, implies that the estimation accuracy 
scales like $L^{-2}$ at the critical points, while it goes like $L^{-1}$ 
at regular points. 
\subsection{Signal-to-noise ratio}
Notice that, in assessing the estimability of a parameter 
$\lambda$, the quantity to be considered is the quantum signal-to-noise ratio 
(QSNR) given by $Q(\lambda)\equiv\lambda^2 G (\lambda)$ which takes into account 
of the scaling of the variance and the mean value of a parameter rather than its 
absolute value. We say that a parameter $\lambda$ is effectively estimable when 
the corresponding $Q(\lambda)$ is large and that to a diverging QFI corresponds 
the optimal estimability. In both cases of few and many spins, at the critical point 
the QFI goes like $1/J^2$, this means that $Q(J)$ is independent on
$J$ and one can estimate large as well as small values of parameters without 
loss of precision.
\section{QET at finite temperature}\label{Tfinite}
In this section we consider the estimation the coupling constant $J$ 
at finite temperature. We first discuss in some detail the short chains 
with $L=2,3,4$ and then we treat the case $L\gg1$.
\subsection{Small $L$}
As a warm-up let us first focus on the simplest, $L=2$ case. A first
step in the computation of the SLD for two qubit is to find the SLD in
the single qubit case. Consider a system with "Hamiltonian"
$H=\mathbf{a} \cdot \s$ in the state $\rho=e^{-\mathbf{a} \cdot \ks}
Z^{-1}$ where $Z=\mathrm{Tr} e^{-\mathbf{a} \cdot \ks}$, and the
three-component vector $\mathbf{a}$ depends on parameter $J$. The SLD
relative to this state turns out to be 
\begin{align}
\Lambda=&-\tanh\left(a\right)\left(\partial_{J}\hat{\mathbf{a}}
\cdot\s\right)\nonumber \\
 &-\left[1+\tanh\left(a\right)-2\tanh\left(a\right)^{2}
 \right]\left(\partial_{J}a\right)\left(\hat{\mathbf{a}}
 \cdot\s\right)\,. \label{sld-1qubit}
\end{align}
where $a$ is the modulus of  $\mathbf{a}$ and $\hat{\mathbf{a}}= \mathbf{a}/a$. 
Now note that the Hamiltonian (\ref{HIsing}) for $L=2$ (with PBC), has the following 
block-diagonal form in the basis $\left\{ |++\rangle,\,|--\rangle,\,|+-\rangle,\,|-+\rangle\right\}$:
\begin{equation}
H=-2\beta\left(\begin{array}{cc}
J\sigma^{x}+h\sigma^{z} & 0\\
0 & J\sigma^{x}\end{array}\right).
\end{equation}
We can then apply formula (\ref{sld-1qubit})  in each subspace to obtain
the full SLD. After some algebra one realizes that the SLD has the
following form
\begin{equation}
\Lambda=c_{1}\sigma^{x}\otimes\sigma^{x}+c_{2}\sigma^{y}\otimes\sigma^{y}+c_{3}
\left(\sigma^{z}\otimes\1+\1\otimes\sigma^{z}\right),
\end{equation}
where $c_{1,2,3}$ are constants which depend on $\beta, J$, and $h$.
When the temperature is sent to zero the above expression becomes
\begin{align}
\Lambda_{T=0}=&\frac{h}{2\left(J^{2}+h^{2}\right)^{3/2}}
\Big[h\left(\sigma^{x}\otimes\sigma^{x} -\sigma^{y}\otimes
\sigma^{y}\right)\nonumber \\
&-J\left(\sigma^{z}\otimes\1+\1\otimes\sigma^{z}\right)
\Big].\label{SLD2}
\end{align}
We see that, already in the simple two-qubit case, the SLD is a 
complicated operator both at positive and at zero temperature. 
More involved expressions are obtained $L=3,4$ and larger.
We do not report here the analytic expression of the corresponding
QFIs $G_J$ for $L=2,3,4$ since they are a bit involved. 
Rather, in order to assess estimation precision at finite 
temperature and compare it to that at $T=0$, 
we consider the ratio $\gamma_J=G_J(\beta,J,h)/G_J(\infty,J,h)$, for some fixed
values of $J$  and illustrate its behavior
by means of few plots. \par
\begin{figure}[h]
\includegraphics[width=0.22\textwidth]{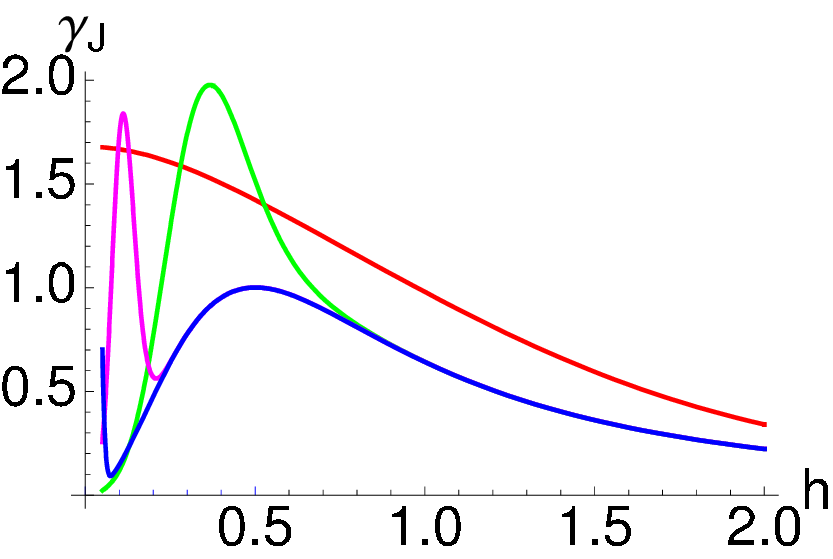}
\includegraphics[width=0.22\textwidth]{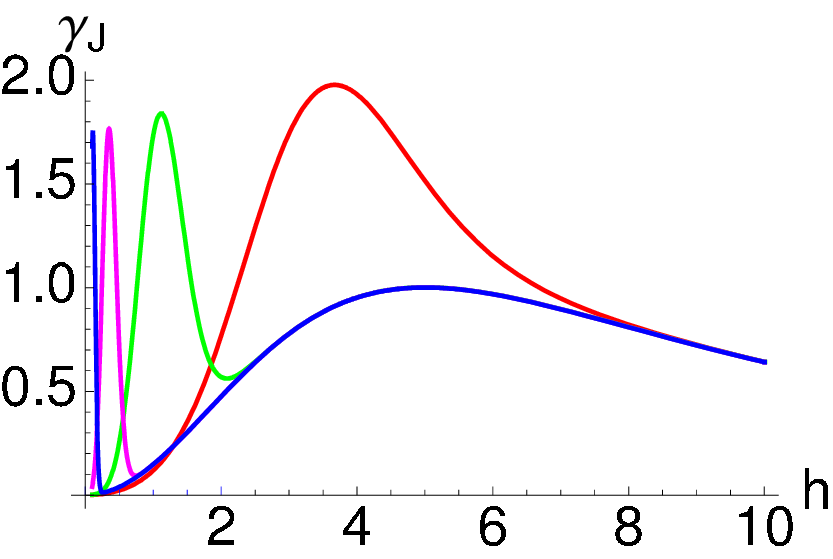}
\includegraphics[width=0.22\textwidth]{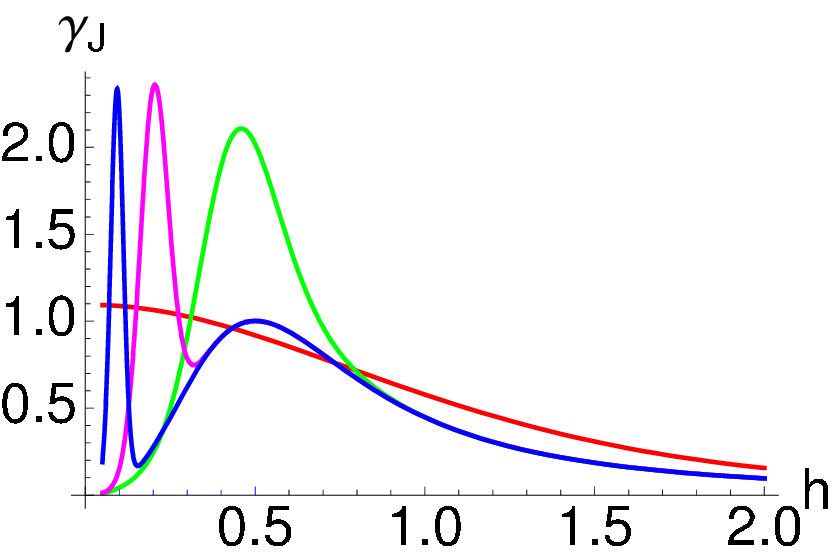}
\includegraphics[width=0.22\textwidth]{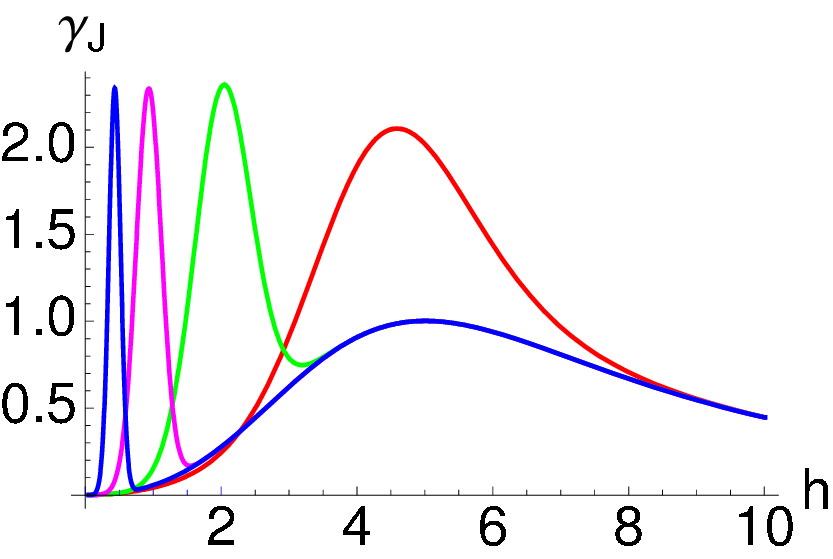}
\includegraphics[width=0.22\textwidth]{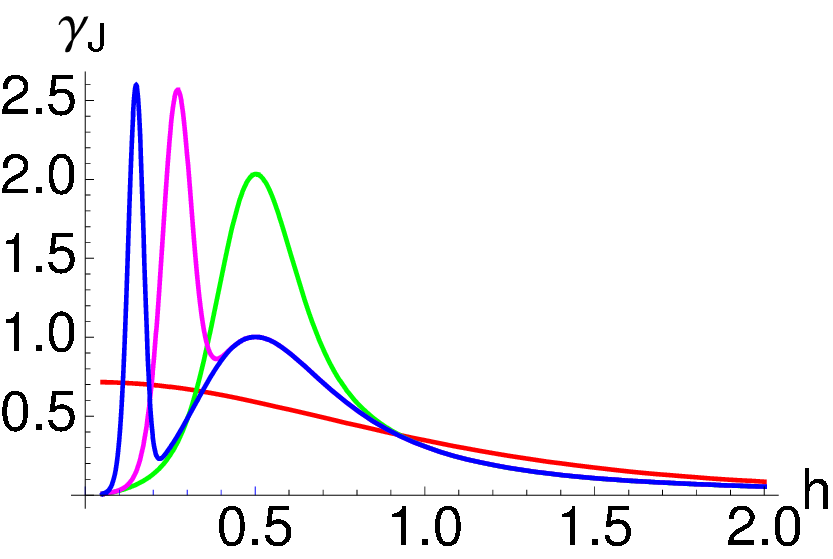}
\includegraphics[width=0.22\textwidth]{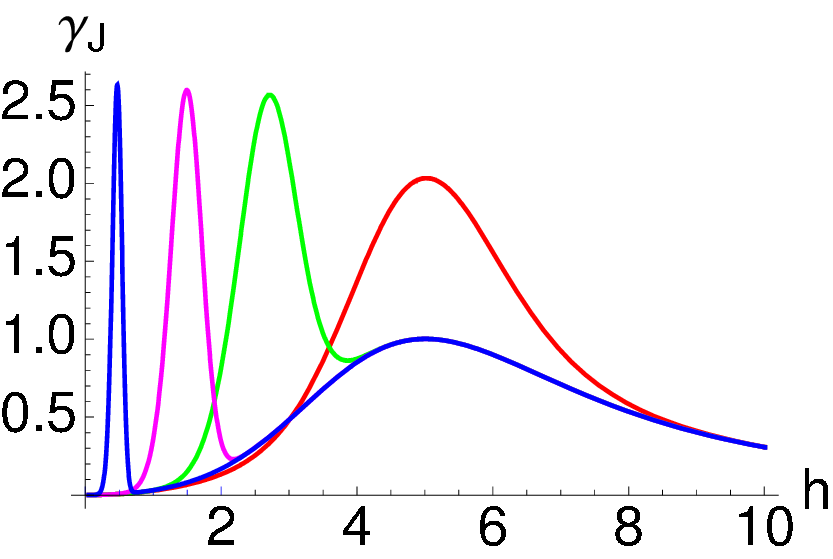}
\caption{The ratio $\gamma_J$ as a function 
of $h$ for $L=2,3,4$ (from top to bottom) and 
$J=0.5$ (on the left), $J=5$ (on the right). The colors 
refer  to different values of $\beta=1$ (red), 
$\beta=10$ (green), $\beta=100$ (purple), $\beta=1000$ 
(blue).} \label{QFI234}
\end{figure}
\begin{figure}[h]
\includegraphics[width=0.23\textwidth]{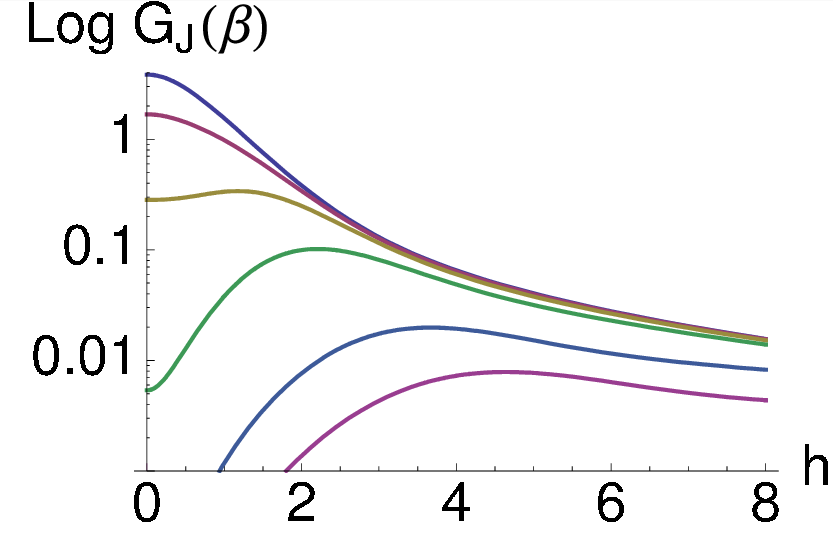}
\includegraphics[width=0.23\textwidth]{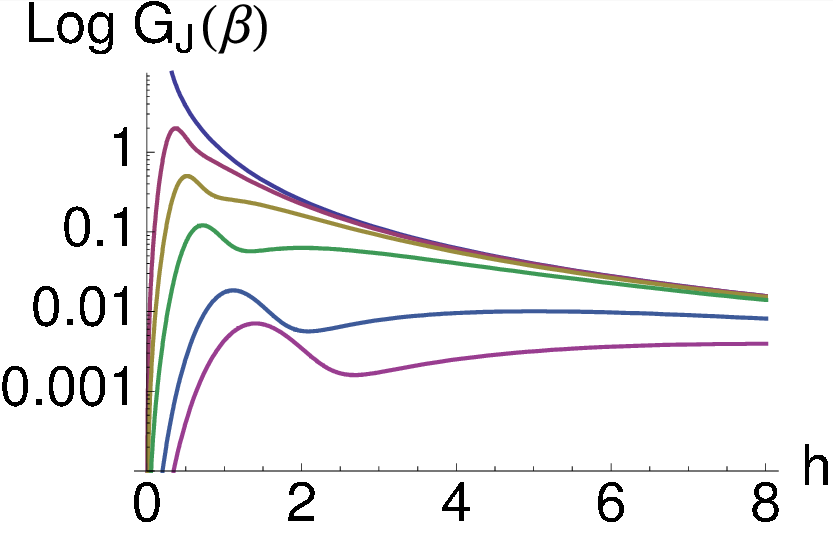}
\caption{Logarithmic plot of $G_J (\beta)$ vs $h$ for 
(from top to bottom) $J=0.1$, $0.5$, $1$, $2$,$5$, $8$ and 
$\beta=1$ (left), $\beta=10$ (right).} \label{Glogaritmo}
 \end{figure}

As it is apparent from Fig. \ref{QFI234} for small $h$ the ratio 
is smaller than $1$, {\em i.e} estimation of $J$ is more precise
at zero temperature, whereas for increasing $h$ a finite temperature may
be preferable. In turn, for any value of $J$ and $\beta$, there is a field
value that makes finite temperature convenient: this is true also 
for low temperature as proved by the presence of a global 
maximum for small  $h$, besides the local maximum at $h=J$. 
For $\beta \rightarrow \infty$ the 
maxima at small $h$ disappears and we recover the zero temperature 
results. Notice that, in view of Eqs. (\ref{gtz}), the ratio $\gamma_J$
is proportional to the QSNR.
Besides, since the 
maxima of $\gamma_J$ vary with $\beta$, we conclude that the optimal 
field $h^\ast$ which maximizes $G_J(\beta)$ varies with the temperature. 
This is illustrated for $L=2$ in Fig. \ref{Glogaritmo}, where we report 
the log-linear plot of $G_J (\beta)$ as a function of $h$ for different 
values of $J$ and $\beta$. For high temperature the maxima 
are localte at a field value close to zero, whereas for decreasing 
temperature they move towards $h^\ast=J$.
\subsection{Large $L$}
At positive temperature and $L$ large, the sums in equation
(\ref{g-QFI}) may be replaced by $L \int dk$. The quantity  $\tilde{G}_J
\equiv G_J/L$ is always convergent, and the convergence rate is
exponentialy fast in $L$ in the (renormalized classical) region $T\ll
\Delta$ whereas is effectively only algebraic when $T\gg \Delta$ (the
quantum-critical region).  Thus, up to contribution vanishing with $L$,
$\tilde{G}_J=\tilde{G}_J^1+\tilde{G}_J^2$ is a bounded function of its
arguments as long as $T>0$, given by 
\begin{eqnarray}
\tilde{G}_J^1 &=& \frac{\beta^{2}}{8\pi}\int_{0}^{\pi}
\frac{dk}{\cosh^{2}\left(\beta\Lambda_{k}/2\right)}\frac{
\left(J+h\cos\left(k\right)\right)^{2}}{\Lambda_{k}^{2}}\\
\tilde{G}_J^2 &=&
\frac{1}{2\pi}\int_{0}^{\pi} dk
\frac{\cosh\left(\beta\Lambda_{k}\right)-1}{\cosh\left(
\beta\Lambda_{k}\right)}\frac{h^{2}\sin\left(k\right)^{2}}{\Lambda_{k}^{4}} \,.
\end{eqnarray}
For any $T>0$ the function $\tilde{G}_J$ has a cusp in $h=J$, where it achieves 
its maximum value. Changing variable from momentum to energy, the integrals 
above can be approximately evaluated in the quantum critical region $\beta
\left|J-h\right|\ll1$ (actually we also require low temperature, i.e. 
$\beta\left|J+h\right|\gg1$). The result is
\begin{eqnarray}
\tilde{G}_{J}^1&=&
\frac{9\zeta\left(3\right)}{8\pi}\frac{T}{J^{2}\left|J+h\right|}+
O\left(T^{0}\right) \\
\tilde{G}_{J}^2&=&
\frac{\mathcal{C}}{\pi^{2}}\frac{\left|J+h\right|}{TJ^{2}}-
\frac{1}{8J^{2}}+O\left(T\right)  \, ,
\end{eqnarray}
where $\mathcal{C}$ is Catalan's constant $\mathcal{C}=0.915$ and 
the Riemann Zeta-function gives $\zeta (3) = 1.202$.  
\par 
In summary, for large sizes and at positive temperature, the maximum of the 
QFI as a function of $h$ is always located at $h=J$ for all values of $J, T$. 
At the maximum, the QFI is approximately given by
\begin{equation}
G_J \simeq \frac{2\mathcal{C}}{\pi^{2}}\frac{L}{TJ}.
\end{equation}
As a consequence, the QSNR scales as $Q_J \sim JL/T$, in other words, 
at finite temperature, the estimation of small values of the coupling 
constant is unavoidably less precise than the estimation of large values. 
As expected, large $L$ and/or low temperature improve the precision of estimation. 
\section{Practical implementations}\label{magnetization}
The SLD represents an optimal measurement, {\em i.e.} the corresponding 
Fisher information is equal to the QFI. However, as we have seen [see 
e.g.~Eq.~(\ref{SLD2})], generally the SLD does not correspond to an observable 
whose measurement can be easily implemented in practice. Therefore, in this 
section, we consider the total magnetization $M_z=\frac{1}{L}\sum_i\sigma^z_i$, 
as a feasible and natural measurement to be performed on the system in order 
to estimate the coupling $J$. We assume that the system is at thermal equilibrium, 
$\rho=Z^{-1} e^{-\beta H}$, and consider short chains $L=2,3,4$. We illustrate 
the procedure in detail for the simplest $L=2$ case. Upon measuring $M_z$, the 
possible outcomes are $m=\{ 1,0,-1\}$ with eigenprojectors $P_m$ given by
$P_1 =\ketbra{00}{00}$, $P_{-1} =\ketbra{11}{11}$, and
$P_0 =\ketbra{10}{10} + \ketbra{01}{01}$.
The  corresponding probabilities $p(m |J)=\Tr (\rho P_m)$ are given by
\begin{align}
p (\pm 1 | J) =&\frac{ \cosh (2\beta\sqrt{J^2 + h^2})}{2 \left [ 
\cosh (2\beta J) + \cosh (2\beta\sqrt{J^2+h^2}) \right ]} \times  \\
  & \Big( 1 \pm h (J^2+h^2)^{-1/2}\tanh (2\beta\sqrt{J^2+h^2})\Big ) 
\nonumber  \\ p(0 | J) =&\frac{\cosh (2\beta J)}{\cosh (2\beta J )+ 
\cosh (2\beta\sqrt{J^2 + h^2})}\,.
\end{align}
The FI is then obtained by substituting $p (m| J)$ into 
Eq.~(\ref{fisherclassica}). The resulting expression provides a bound 
for the variance of any estimator of $J$  based on $M$ measurements 
of magnetization: $\hbox{Var} (J) \geq 1/ M F_J$.
\begin{figure}[h]
\includegraphics[width=0.4\textwidth]{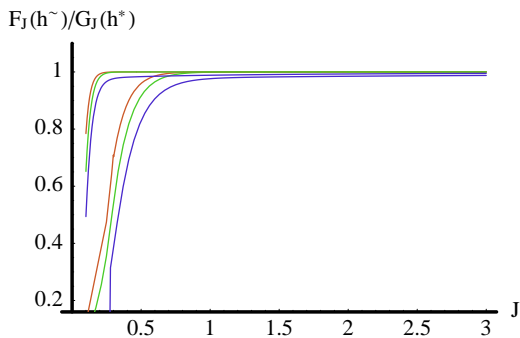}
\caption{The ratio $F_J (\beta, J, h\hbox{\~{}})/G_J(\beta, J,h^\ast)$ 
as a function of $J$ for $L=2$ (red line), $L=3$ (green line), and 
$L=4$ (blue line). The bottom group of lines is for $\beta=3$, whereas
the top group is for $\beta =10$.} 
\label{FJGJ23}
\end{figure}\par
The Braunstein-Caves inequality says that the FI of any 
measurement $F_J$ is upper bounded by the QFI $G_J$. For the
magnetization this is illustrated in Fig. 
\ref{FJGJ23}, where we plot of the ratio $F_J (\beta, J, h\hbox{\~{}})/
G_J(\beta, J,h^\ast)$ for $L=2,3,4$, $h\hbox{\~{}}$ being the field maximizing
the FI. Notice
that for increasing $J$ the FI of the magnetization saturate to the QFI,
{\em i.e.} magnetization measurement becomes optimal. The saturation is
faster for lower temperature (we report the ratio for $\beta=3$ and
$\beta=10$). Notice also that for low temperature the dependence of 
the ratio on the size $L$ almost disappears. In summary: for any temperature
there is a threshold value for $J$, above which the measurement
of the magnetization is optimal for the estimation of $J$ itself.
This threshold value decreases with temperature, and for zero
temperature magnetization is optimal for any $J$.
Indeed, after explicit calculation of 
Eq.~(\ref{fisherclassica}) for $L=2,3,4$ 
we found that, in the limit $T\to 0$, $F_J (h,T=0)=G_J(h,T=0)$, i.e.~
the FI of the magnetization is equal to the QFI. In other words the
estimation based on magnetization may achieve the ultimate 
bound to precision imposed by quantum mechanics. Besides, at finite temperature, 
despite the fact that the equality does not hold exactly, $F_J$ is only 
slightly grater than $G_J$ almost in the whole parameter range $(J,T)$. 
This may be also seen in the behavior of $F_J$ versus temperature: 
the ratio $\delta_J = F_J (\beta, J,h)/F_J  (\infty, J, h)$ at fixed $J$ 
may be greater than $1$ for some values of the magnetic field, namely 
magnetization measurements may be more precise at finite $T$, as it 
happens for the optimal measurement with precision bounded by 
the QFI. Of course, for  $T\rightarrow 0$, $\delta_J \to 1$.
\par
\begin{figure}
\includegraphics[width=0.4\textwidth]{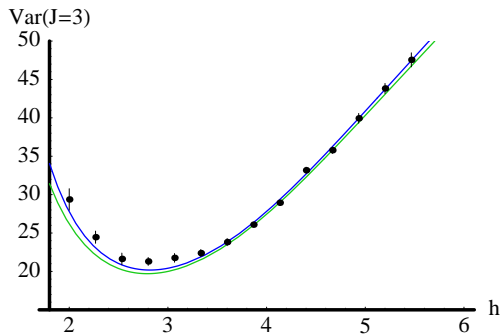}
\caption{Estimation of the coupling constant by magnetization
measurement and Bayesian analysis. The plot is for $L=2$ and $\beta=1$
with true value of the coupling equal to $J=3$. We report the variance
of the Bayesian estimator for Monte Carlo simulated experiments (M=500,
black dots), the corresponding variance evaluated using the asymptotic
a-posteriori distribution (blue line), and the (classical) 
Cramer-Rao bound (green line).}\label{Bayes}
\end{figure}

Overall, we conclude that the magnetization $M_z$ is a good candidate 
for nearly optimal estimation. Of course we still need an efficient 
estimator, that is an estimator actually saturating the (classical) 
Cramer-Rao bound. To this aim we employ a Bayesian analysis, since 
Bayes estimators are known to be asymptotically efficient \cite{LeCam}, {\em i.e.} 
$\hbox{Var} (J_B) = 1/M F_J$, for $M\gg 1$, $J_B$ being the Bayesian 
estimator (see below).
According to the Bayes rule, given a set of outcomes 
$\{ m \}$ from $M$ independent measurements of the magnetization, 
the a-posteriori distribution for the parameter $J$ is given by
$p(J | \{ m \}) = 1/N\prod_{m} p(m| J)^{n_m}$
where $N$ is a normalization constant and $n_m$ is the number of 
measurements with outcome $m$. Bayes estimator is the mean 
$J_B = \int dJ\:J\: p(J | \{ m \})$ of the 
a-posteriori distribution and precision is quantified by the
corresponding variance. In the asymptotic limit of many measurements 
$M\gg 1$,  
$n_m \to M p(m |J^\ast)$ where $J^\ast$ is the true value of the 
parameter to be estimated and the a posteriori distribution
rewrites as 
$p_a(J | \{ m \}) = 1/N
\sum_m \exp \left[ M p(m |J^\ast)\log p(m |J)\right]$.
In order to check the actual meaning of 
"asymptotic" we have performed a set of Monte Carlo simulated experiments 
of the whole measurement process.
In Fig. \ref{Bayes}, we report the result of Monte Carlo simulated
experiments of magnetization measurements for $J=3$ and $\beta=1$. 
The black dots represent the mean variance of the Bayes estimator $J_B$
averaged on $20$ sets each of $500$ measurements.      
The blue line is the corresponding variance evaluated using the
asymptotic a-posteriori distribution, whereas the green line is 
the Cramer-Rao bound $(MF_J)^{-1}$. 
The plot shows that Bayes estimator is indeed asymptotically 
efficient and that already with a few hundreds of 
measurements one may achieve the ultimate precision.
Overall, putting this result together with the fact $F_J \simeq G_J$ 
(see Fig. \ref{FJGJ23}) we conclude that the measurement of 
the total magnetization of the system provides a nearly optimal
and feasible measurement (at any $\beta$) to estimate the coupling 
of the one-dimensional quantum Ising model.
\section{Conclusions}\label{conclusion}
The coupling constant of a many-body Hamiltonian is not an observable
quantity and we have to solve a quantum statistical model to evaluate
the bounds to its estimation precision. This has fundamental
implications since it corresponds to find the ultimate limits imposed
by quantum mechanics to the distinguishability of different states of
matter. In this paper we exploited the equivalence between the quantum
Fisher metric and the (ground or thermal) Bures metric and all the
results recently obtained for the latter.  Specifically at zero
temperature, the Bures metric scales with the system size $L$ at regular
points whereas it can increases as $L^2$ at or in the vicinity of
quantum critical point. A similar enhancement takes place when
temperature is considered. In turn it is possible to exploit this
enhancement to  dramatically improve the bounds to precision in a quantum
estimation problem.  Let us imagine that an experimenter would like to
infer the value of a coupling constant of a physical system over which
he has little or no control. Reasonably the experimenter has good
control over the external fields he can apply to the system. The idea is
then to tune the external field to a value close to the quantum critical
point. At this value of the couplings, an improvement of order of $L$
can be achieved in the precision of the estimation of the unknown
coupling.  To test these ideas in practice, we have worked out in detail
a specific example, the 1D quantum Ising model. This model provides us
with all the ingredients we need, a coupling constant $J$, an external
field $h$, and a quantum critical point at $h=J$.  The main
accomplishments of our analysis are:  i) At zero temperature we
evaluated the precision in  the estimation of the coupling, exactly for
short chains of $L=2,3,4$ sites and asymptotically for large $L$. We
found that the optimal estimation is possible at values of the field
exactly equal to the critical point, independently of $L$. For large $L$
we indeed observe a $1/L$ enhancement of precision, and a quantum
signal-to-noise ratio independent of the coupling. ii) At positive
temperature the optimal value of the field is again given by the
critical value when the system size is large or the temperature is low.
In the other working regimes the optimal field maximizing the quantum
Fisher information, defines a set of pseudo-critical points, and 
the optimal precision scales as $TJ/L$. iii) We obtained the
optimal observable for estimation in terms of the symmetric logarithmic
derivative and showed that already in the case $L=2$ it does not
correspond to an easily implementable measurement. iv) We have shown
that for small $L$ the measurement of the total magnetization allows 
to achieve ultimate
precision. Using Monte Carlo simulated experiments and Bayesian analysis
we proved that this is possible already after a limited number of
measurements of the order of few hundreds. We conjecture that this may
be true for any $L$; work along this line in progress.
\par
Overall, we found that criticality is a resource for precise
characterization of interacting quantum systems (e.g.~a quantum
register), and may represent a relevant tool for the development of
integrated quantum networks.
\section*{Acknowledgments}
The authors thank M. G. Genoni, P. Giorda, A. Monras, S. Olivares, and 
P. Zanardi for useful discussions.

\end{document}